# A Distributed Multimedia Communication System and its Applications to E-Learning[*]

Hans L. Cycon, Thomas C. Schmidt, Matthias Wählisch,

Mark Palkow and Henrik Regensburg

**Abstract** — In this paper we report on a multimedia communication system including a VCoIP (Video Conferencing over IP) software with a distributed architecture and its applications for teaching scenarios. It is a simple, ready-to-use scheme for distributed presenting, recording and streaming multimedia content. We also introduce and investigate concepts and experiments to IPv6 user and session mobility, with the special focus on real-time video group communication.

## I. Introduction

A paradigm shift is underway in the Internet. Networked devices, formerly situated on the desks of scientist and business, are now consumer parts and provide information, communication and entertainment. Visual devices performing synchronous communication such as voice or videoconferencing over IP (VoIP/VCoIP) are now ubiquitous and raise new challenges for the Internet infrastructure.

Videoconferencing is about to become a lightweight day-to-day application. This trend follows from high bandwidth data connections, which are increasingly available to the public at reasonable prices. There is also some remarkable progress in video/audio compression algorithms, which reduce the video data stream to less than 1% at the sender site and reconstruct it again to a high quality video sequence on the receiver site. A third reason is that more and more desktop video conferencing software is running on ordinary desktop PCs or laptops, seamlessly using the Internet connections available.

The increasing video/audio quality and the easy-to-use user interface combined with all kinds of application sharing will soon convince more and more users to enrich their communication habits by video components.

This trend is also found in teaching and learning scenarios. There is a rising number of Internet/Intranet-backed courses at universities or private companies. Here also the mail and voice communication will be supplemented by synchronous and asynchronous video components.

As a second trend wireless Internet conquers the airwaves. The availability of new, truly mobile IP enabled sub network layers not only offers connectivity to nomadic users at roaming devices, preserving communication sessions beyond IP subnet changes, but re-raises questions concerning the quality of IP service: The constant bit rate scenarios of voice and videoconferencing will appear significantly disturbed by packet loss intervals, delays or jitter exceeding 100 ms. Thus, when heading towards VCoIP as a standard Internet service, important steps for global usability have to be taken, focusing on ease and quality.

In the present paper we report on a multimedia communication system including a VCoIP software with a distributed architecture and its applications for teaching scenarios. We also introduce and investigate concepts and experiments to IPv6 user and session mobility with the special focus on real-time multicast group communication.

This paper is organised as follows. Section II presents the basic video conference software and its applications in E-learning scenarios. In Section III we discuss some IP-mobility aspects arising in real-time group communications and introduce a multicast solution within the Mobile IPv6. Finally, section IV is dedicated to conclusions and outlook.

## II. A Distributed Communcation System

### A. The Basic Software

The digital audio-visual conferencing system we use is a server-less multipoint video conferencing software without MCU developed by the authors [2]. It has been designed in a peer-to-peer model as a lightweight Internet conferencing tool aimed at email level use. Guided by the latter principle, it refrained from implementing H.323 client requirements [1].

The system is built instead upon a fast, highly efficient video codec, based on a H.264/AVC [3] standard implementation, optimized for low bit rate video streams. The encoder of our implementation (called DAVC) performs up to 1.5 dB PSNR better than the verification model VM 8.2 at a bit rate of 300 kbit/s. At this bit rate the codec runs at a speed of 65 frames per second encoding *and* decoding simultaneously on a Pentium 2.2 GHz processor machine. Note for higher bit rate however the verification model performs better in rate-

[*] This work was supported in part by the EFRE program of the European Commission and by the German Bundesministerium für Bildung und Forschung.

Hans L. Cycon is with the FB 1, FHTW Berlin, Treskowalle 8, 10318 Berlin, Germany (e-mail: hcycon@fhtw-berlin.de).

Thomas C. Schmidt is with the Department of Electrical Engineering and Computer Science, HAW Hamburg, Berliner Tor 7, 20099 Hamburg. He is also with the Computer Center, FHTW Berlin, Treskowallee 8, 10318 Berlin, Germany (e-mail: schmidt@fhtw-berlin.de).

Matthias Wählisch is with the Computer Center, FHTW Berlin, Treskowallee 8, 10318 Berlin, Germany (e-mail: mw@fhtw-berlin.de).

Mark Palkow is with the daViKo GmbH Berlin, Hoenower Strasse 35/PF 16, 10318 Berlin, Germany (e-mail: palkow@daviko.com).

Henrik Regensburg is with the FB 1, FHTW Berlin, Treskowalle 8, 10318 Berlin, Germany (e-mail: h.r@fhtw-berlin.de).

distortion sense due to the more sophisticated motion compensation technique involved (Fig.1).

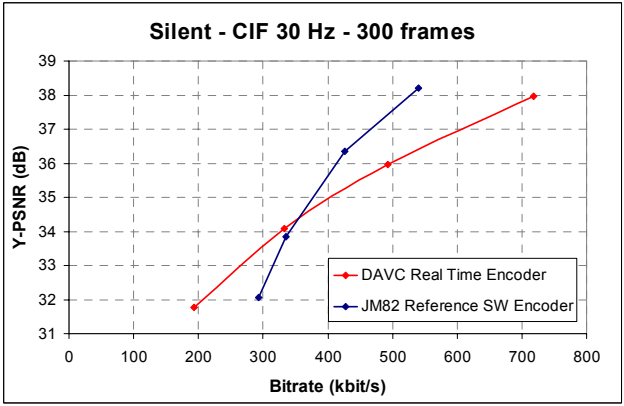

Fig. 1. Rate-Distortion curves comparing H.264/AVC Reference Verification Model 8.2 and proposed encoder algorithm DAVC. Note that for low bit rate there is a distinct advantage of DAVC.

By controlling the coding parameters appropriately, the software permits scaling in bit rate from 24 to 1400 kbit/s on the fly. There are also application-tailored versions using a fast wavelet-based video codec [4] for higher available data rate. Audio data is compressed using a variable bit rate codec [5] with latencies below 120 ms depending on the chosen buffer size.

Audio and video (A/V) -streams can be transmitted by unicast as well as by multicast. An application-sharing facility is included for collaboration and teleteaching. It enables participants to share not only static documents like Power Point files but also any dynamic PC actions like mouse pointer movements or animations. A/V - streams including dynamic application sharing actions can be recorded on any site. These data can be displayed locally or automatically converted into a web streaming format for internet wide availability.

This system is equally well suited to wired and wireless video conferencing on a best effort basis, since the audio/video quality can be controlled to adapt the data stream to the available bandwidth. In faulty network conditions like poor WAN links we use unicast TCP transmissions to avoid distortions. For point-to-multipoint situations like virtual classrooms there is also a possibility to switch to multicast UDP network transmission to minimize computation and transmission load for networks and sender.

For global connection establishment in the Internet we use a dynamic user session recording [6]. We denote this by User Session Locator (USL) and store appropriate session information within an LDAP directory server. The clients update information about ongoing sessions regularly. Our system *restricts* user addressing to email addresses since the only uniformly available user addressing scheme on the Internet is given by mail. A global user look-up proceeds in two steps. Firstly, the mail exchange (MX) record for the target user is requested from the Domain Name System, and secondly, the directory server hostname formed from the above naming convention is resolved (see Fig.2).

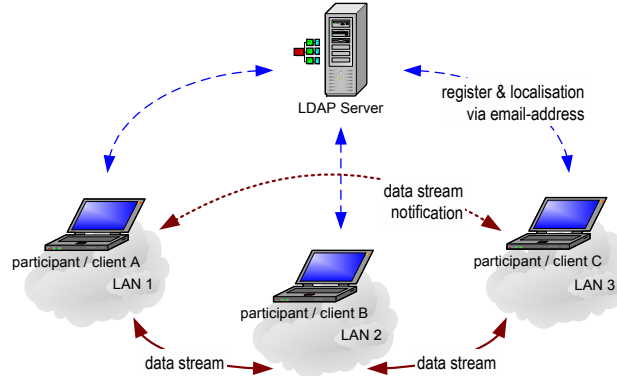

Fig. 2. Connection establishment by registration in an LDAP server and user localization via email address

### B. Distributed Collaborative Teaching Scenarios

The system described above is employed within various learning scenarios:
### 1) Synchronous distributed learning

Teacher and students are connected by intranet. This can be realized by LAN or WLAN. All participants establish mutual connections via web server. The teacher can than send his PC applications including all pointer movements to the students PCs. Students outside the lecture room can participate active and passive by real-time audio/video transmission to there PCs. Since the latencies in both directions are well below 100 ms, real-time conferencing operates smoothly. The teacher can send his talks and presentations - including dynamic applications - to the students via multicast or unicast streams. Students can see and listen, listen only or participate via a video/audio back channel. (See C. for realizations)

Students can also initiate co-operation in small groups via full video conferencing. Within the peer to peer network each student can send, receive and work on any PC-applications for collaboration (Fig.3).

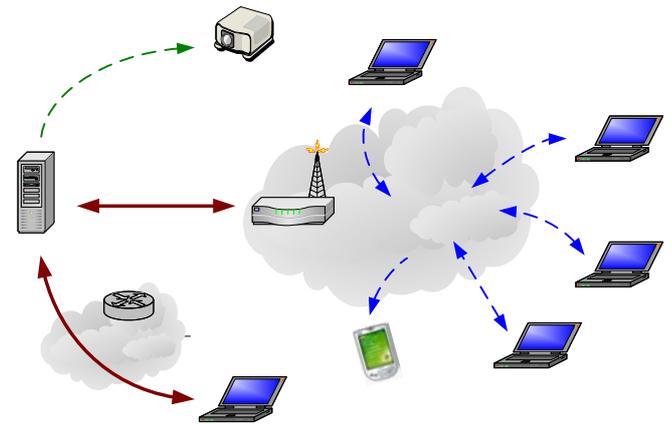

Fig. 3. Distributed WLAN/LAN teaching & presenting scenario

### 2) Asynchronous distributed learning.

Based on our system's capabilities, new scenarios for asynchronous distributed learning evolve. Each station can record all sessions. The data will be stored locally and can be played back in a proprietary format. To make recordings world-wide accessible, it can be converted and uploaded onto a server for streaming, e.g. into a MS streaming format Classroom presentation or distributed group work are thus ready to be played back anywhere at any time (Fig.4).

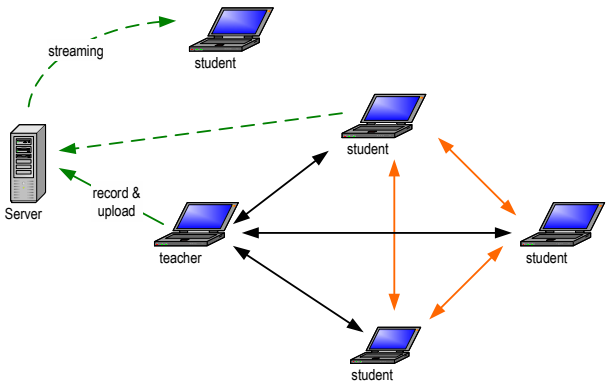

**Fig. 4. Asynchronous/ synchronous distributed learning Scenario**

*C. Realizations*

One version of the "frontal" scenario described in B.1) has been realized at the medical university Charité Berlin in the new histology classroom, where a digital PAL size video signal is send directly from the teacher's microscope to the students PCs by multicast enabled LAN [9]. This replaces a traditional overhead projection device, allowing a significantly enhanced visual presentation at the student's desk. There are 64 student PCs in one classroom installed but there is a planned extension to 120 PCs.

The above described video conferencing technique has also been used in a notebook university project (Medu-Mobile), at Charité Berlin which was designed and realized by T.N. Nguyen-Dobinsky and G. Kaiser [10].

This project was designed to realize and endorse medical bed side teaching scenarios. Patients will be visited only by a camera team instead of a complete group of students. Bed side teaching will be transmitted via WLAN video conferencing live to the students, who can be anywhere on the campus. The students can participate interactively by asking questions to the patient or to the lecturer. In addition to the bed side video medical images (e.g. X-ray or ultrasonic) can be send directly out of medical devices or multimedia data bases to the study group.

The system is currently also in operation for Internet based cross teaching between different universities [8]. On both sites there is a teacher in a classroom with students. The video, combined with dynamic applications of the other venue, is projected at each others site, respectively. The teachers may give their talks alternatively or in a discussion session. All Students can also participate by wireless microphone. All sessions can be recorded and stored at any site in a streaming format. Supplementary, the technique has been used for taking examinations via the Internet, for example between Berlin and Chicago.

A fast broadband conference version of our system to connect two distant medical campus´ is on schedule. It works with two beamers at one side, to project the full size video of the speaker and his presentations including highly dynamic animations. The speaker's video as well as the presentation is digitized by high performance XGA frame grabbers. The video stream will be encoded with a very fast software wavelet-based codec. This includes also a special application sharing device using a H.264 standard intra codec to transmit screen applications.

### III. SESSION MOBILITY & GROUP COMMUNICATION

*A. Changing Networks*

When users move around while operating their Internet devices (laptops, PDAs etc.), we have to consider changes of IP subnets during conferencing sessions. As it is common standard in mobile phones, preserving VCoIP sessions beyond renumbering is compulsory. The next generation Internet grants the ability of coping with multiple and changing addresses, thus giving rise to a seamless migration of connected devices while running in service.

The fundamental approach to Internet mobility is the recently appointed Mobile IPv6 (MIPv6) Internet Standard [7]. MIPv6 transparently operates address changes on the IP layer as a device moves from one network to the other by sustaining initial IP addresses and hiding the different routes to the socket layer. In this way hosts are enabled to maintain transport and higher-layer connections, when they change locations.

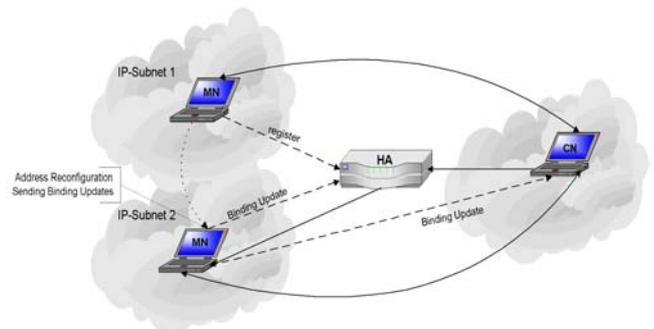

**Fig. 5. Mobile Client in IPv6**

*B. VCoIP in MIPv6*

Real-time video communication imposes stern quality of service requirements on the underlying network infrastructure. 100 ms real-time carry relevant information, a spoken syllable for instance in the audio case. More generally network disturbances exceeding 300 ms interrupt a video conference at the user's level, whereas perturbations lasting less than 100 ms remain tolerable and may even be adjusted by jitter-hiding buffers. In mobile IPv6 the timing of handover procedures consequently forms a critical issue: In entering a new IP network, i.e. after completing the layer 2 handoff, the Mobile

Node (MN) instantaneously has to perform an automatic address reconfiguration followed by binding updates with its Home Agent (HA) and the Correspondent Node (s. Fig. 5). During this handover procedure the MN is unable to communicate until the HA has learned its new Care-of Address. Packets may then proceed through the HA as a forwarder with the likely result of increased delay and jitter. The mobile IPv6 handover procedures consist of two principal tasks: the local, topology independent handoff and re-addressing, and the distant Binding Updates with strong topology dependence. Temporal optimisation consequently needs a dual focus. The acceleration of local steps mainly improves through layer 2 speed-up and stack implementations optimising the interplay of the layers 2 and 3. Delay hiding proxy techniques such as the Hierarchical Mobile IPv6 [12] are needed to overcome roundtrip delays during Binding Updates.

Challenges are tightened by multicast-based group communication. In conferencing scenarios each member commonly operates as receiver and as sender. A mobile environment therefore needs to cope with the tardy source specific construction of multicast routing trees. Multicasting in a mobile Internet environment poses the fundamental problem that routing information bases need to be built from possibly rapidly changing group member locations.

*C. Mobile Internet Multicasting*

Multicast group communication raises quite distinctive aspects within a mobility aware Internet infrastructure: On the one hand Multicast routing itself supports dynamic route configuration, as members may join and leave ongoing group communication over time. On the other hand multicast group membership management and routing procedures are intricate and too slow to function smoothly for mobile conference users. In addition multicast imposes a special focus on source addresses. Applications commonly identify contributing streams through source addresses, which must not change during sessions, and routing paths in most protocols are chosen from destination to source.

In general the roles of multicast senders and receivers are quite distinct. While a client initiates a local multicast tree branch, the source may form the root of an entire source tree. Hence multicast mobility at the sender side poses the more delicate problem. Multicast support in Mobile IPv6 avoids all these challenges by bi-directional tunnelling multicast packets through the Home Agent. Mobility in this approach is completely hidden from routing at the price of triangular paths, largely increasing delays and jitter. Within this section we will introduce our proposal to seamlessly integrate multicast suitable for real-time applications in a mobile Internet infrastructure.

"Seamless Multicast Handovers in a Hierarchical Mobile IPv6 Environment (M-HMIPv6)" [11] extends the Hierarchical MIPv6 [12] architecture to support mobile multicast receivers and sources. Mobility Anchor Points (MAPs) as in HMIPv6 act as proxy Home Agents, controlling group membership for multicast listeners and issuing traffic to the network in place of mobile senders. Note that HMIPv6 concepts request MAPs to be in close neighbourhood of the Mobile Node with the result of negligible intercommunication delays.

All multicast traffic between the Mobile Node and its associated MAP is tunnelled through the access network. Handovers within a MAP domain remain invisible in this micro mobility approach. In case of an inter–MAP handover, the previous anchor point will be triggered by a reactive Binding Update and act as a proxy forwarder. A Home Address Destination Option, bare of Binding Cache verification at the Correspondent Node, has been added to streams from a mobile sender. Consequently transparent source addressing is provided to the socket layer. Bi-casting is used to minimize packet loss, while the MN roams from its previous MAP to a new affiliation (s. Fig. 6). A multicast advertisement flag extends the HMIPv6 signalling. In cases of rapid movement or crossings of multicast unaware domains, the mobile device remains with its previously associated MAP. Given the role of MAPs as Home Agent proxies, the M-HMIPv6 approach may me viewed as a smooth extension of bi-directional tunnelling through the Home Agent supported in basic MIPv6.

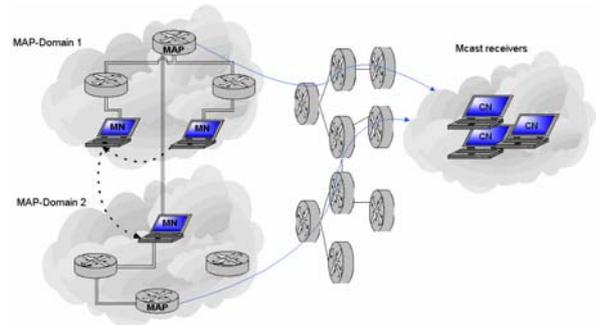

Fig. 6. Mobile Multicast Source

The M-HMIPv6 multicast mobility scheme has been evaluated with respect to its handover performance, its handover frequency, its robustness and its protocol overheads [13]. It was found that reactive handovers as employed in M-HMIPv6 admit a reasonable time to completion of about 75 ms, comparable to predictive handover techniques around. Reactive handovers at the same time operate with little overhead and reliable robustness. From a quantitative study of expected handover occurrences it could be shown that the presence of Mobility Anchor Points significantly reduces handover frequencies. This smoothing effect gains additional importance by observing instability of techniques using predictive handovers in the case of Mobile Node's rapid movement.

## IV. CONCLUSION & OUTLOOK

In this paper we present a distributed communication framework and conferencing software and some applications to e-learning scenarios. The applications include an easy-to-use scheme for distributed presenting, recording and streaming of multimedia content. The video conferencing module is based on a H.264/AVC software implementation. In addition to this we use also some costumer tailored wavelet-based codecs.

A simple scheme, compliant with current Internet infrastructure, for locating nomadic users at roaming sessions forms also part of our conferencing system.

As a second key issue we consider Internet mobility under real-time communications. Even though a principle feasibility of IP mobility for real-time video communication could be demonstrated, MIPv6 handover procedures need tightening. Future improvements need to focus on a reduction of packet loss probabilities as seem attainable by reviewing Mobile Node stack properties and appropriate buffering opportunities.

In the context of group conferencing scenarios, multicast solutions are vital to Internet mobility infrastructure. We presented an approach for seamless integration of real-time multicast mobility and briefly sketched performance enhancements gained from this solution.

**Hans L. Cycon** is currently teaching mathematics and signal processing at FHTW Berlin, University of Applied Sciences. He received his diploma in physics in 1975 and his PhD in mathematics in 1979 and his habilitation in 1984 from the Technical University Berlin, Germany. His publications fields are mathematical physics and signal processing i.e. image coding. Hans L. Cycon is leading several projects in developing wavelet based still image and video compression codecs. He is member of the German delegation of the ITU/ISO standardization committee for JPEG 2000 still image standard.

**Thomas C. Schmidt** is teacher of Information Engineering at the HAW Hamburg and project manager at FHTW Berlin, where he was head of the computer centre for many years.
He studied mathematics and physics at Freie Universität Berlin and University of Maryland, USA. In 1993 he received his PhD in mathematical physics for a work in many particle theory of quantum mechanics done at the theory group of the Hahn-Meitner-Institut Berlin. Since the late 1980s he has been involved in many computing projects, especially focusing on simulation and parallel programming, distributed information systems and visualisation. His current fields of interest lie in the areas of mobile and multimedia networking and hypermedia information processing, where he has continuously conducted numerous projects on national and international level.

**Matthias Wählisch** is a member of the networking group of the computer centre of FHTW Berlin. He is studying mathematics and computer science at Freie Universität Berlin. His major fields of interest lie in networking protocols, where he looks back on five years of professional experience in project work and publication.

**Mark Palkow** presently is the Managing Director and Chief Developer at the daViKo Gesellschaft für digitale audiovisuelle Kommunikation mbH that he founded in 2000. He received his diploma in communication engineering from the Fachhochschule Telekom Berlin in 1996. Since then he has worked on several research projects at FHTW Berlin, the Old Dominion University Norfolk and the Heinrich Hertz Institut Berlin.

**Henrik Regensburg** is a member of the developer group of the "competence center media and networks" of FHTW Berlin. His major fields of interest lie in distributed video applications and coding, networking and a/v-content authoring. He received his diploma in applied computer science from the University of Applied Sciences FHTW-Berlin in 2002. Since then he has worked on research projects at FHTW Berlin and several freelance projects in commerce, all concerned with video conference technology and e-learning.